\documentstyle[12pt,epsf]{article} 
\begin{document} 
\setcounter{page}{1} 
\parskip=0pt plus2pt 
\begin{center} 
{\bf Exact Results For The Adsorption Of A Semiflexible Copolymer Chain In Three Dimensions}
 
\end{center} 
\begin{center} 
Pramod Kumar Mishra  \\ 

\vspace*{.5cm} 
{\it 
 Department of Physics, DSB Campus\\ 
  Kumaun University, Naini Tal-263 002 \\ 
     Uttarakhand, India } 

{\bf Email: {\it pkmishrabhu@gmail.com}}

\end{center} 
\vspace*{1cm} 
\baselineskip=10pt 
 
\noindent 
Lattice model of directed self avoiding walk has been solved analytically to investigate adsorption
desorption phase transition behaviour of a semiflexible sequential copolymer chain on a two dimensional 
impenetrable surface perpendicular to the preferred direction of the walk of the copolymer chain in three dimensions.
The stiffness of the chain has been accounted by introducing an energy barrier
for each bend in the walk of the copolymer chain. 
Exact value of adsorption desorption 
transition points have been determined using generating function method for the cases in which one type of monomer is having 
interaction with the surface viz., (i) no interaction (ii) attractive interaction and (iii) repulsive interaction. 
Results obtained in each of the case show that for stiffer copolymer chain adsorption transition
occurs at a smaller value of monomer surface attraction than a flexible copolymer chain.  
These features are similar to that of a semi-flexible homopolymer chain adsorption.\bigskip 
\section{Introduction}
Biopolymers ($DNA$ and $Proteins$) are known to exhibit under different physiological conditions a variety 
of persistent lengths ranging from
being much smaller than the over all length of the polymer chain, to being comparable
to the polymer chain length \cite{1,2}. The persistent length
of biopolymers is in between the flexible and stiff polymer chains therefore biopolymers
are said to be semiflexible. In addition to it such polymers are made of different type of monomers. These
monomers are randomly distributed along the polymer chain length. Therefore, biopolymers are
random copolymers. For example, protein molecules are composed of hetrogeneous sequence of hydrophobic
and hydrophilic residues and therefore proteins can be considered as a random copolymer.
The conformational properties of such polymer chains have attracted considerable attention
in past few years because of advancement in the experimental methods in which it has become 
possible to pull and stretch such single bio-molecule to measure its elastic properties 
\cite{3,4}. These study reveal a wealth of information about the conformational 
behaviour of biopolymers and therefore of biological importance.

Interest in copolymer chains are not only due to their application in biological
physics but also due to their other variety of applications. For example, application
of copolymer is in the field of biosensors, 
pattern recongnition based application, 
adhesion, surface protection. In addition to it, study of adsorption of the copolymer chain 
on a surface is also useful in determining 
the relation between their composition with their adsorption characteristics.

Conformational properties of a linear homopolymer chain in dilute solution 
have been extensively studied \cite{5} and its adsorption on surface
has also been a well understood problem \cite{6,7,8,9,10}. In past few years 
sequential copolymer adsorption in two dimension has also received attention \cite{11}
due to location of its adsorption transition point and calculation 
of crossover exponent. The problem of random copolymer adsorption has been extensively studied
using numerical methods; for instance, see, \cite{12,13,14,15,16,17} and references theirin. However, analytical methods for
adsorption of copolymer chain with self avoidance effect are limited to 
directed walk like models \cite{11}. The study of adsorption of
such copolymer chain merits somewhat differently from homopolymer case 
in the sense that different type of monomers of the copolymer chain need not have attractive
interaction with the surface.

The essential physics associated with the behaviour of a surface
interacting polymer chain in a good solvent can be derived
using a model of self-avoiding walk (SAW) on a suitable lattice.
If surface is attractive, it contributes an energy $\epsilon_s$
($<0$) for each step of walk lying on the surface. This leads
to an increased probability defined by a Boltzmann weight
$\omega=\exp(-\epsilon_s/k_BT)$ of moving a step on the surface
($\epsilon_s < 0$ or $\omega > 1$, $T$ is temperature and
$k_B$ is the Boltzmann constant). 
For $\omega< \omega_c$ polymer chain is found in desorbed phase and it gets adsorbed
for $\omega>\omega_c$.
The polymer
chain gets adsorbed on the surface at a suitable value of $\epsilon_s$. 
The transition
between these two regimes (adsorbed and desorbed) is marked by a critical value of adsorption
energy or $\omega_c$. 
One may define the crossover
exponent $\phi$ at the adsorption transition point as, $N_s \sim N^{\phi}$, where $N$ is the total
number of monomers in the polymer chain and $N_s$ the number of monomers adsorbed on the surface. 

In this paper, we have extended lattice model of directed self avoiding walk introduced
by Privman {\it et al.} \cite{18} for homopolymer chain, to study the 
adsorption properties of a semiflexible copolymer chain 
immersed in a good solvent in three dimensions
and also to examine the question whether such 
copolymers differ from homopolymers with respect to their critical behaviour.
However, directed walk model has been used by Privman {\it et al.} \cite{18} 
to describe rod-coil transition of a linear polymer chain and analysing the appropriate
scaling properties of the polymer chain in rod-coil transitions. Mishra {\it et al.} \cite{19} 
used definition of Privman {\it et al.} \cite{18} of directed walk model to study adsorption behavior of the 
semiflexible polymer chain.

We have considered semiflexible sequential copolymer chain composed of two type of monomers (A \& B).
Such copolymer model serve as a paragmatic model for actually disordered macromolecules
(for example, proteins).
For the adsorption of semiflexible sequential copolymer chain on an impenetrable surface 
perpendicular to the preferred direction of the chain we have solved the 
directed walk model analytically and have found exact critical value of the surface attraction
for the adsorption of the copolymer chain in three dimensions. 

The outline of the paper is as follows: In Sec. 2 we describe the lattice model of
directed self avoiding walk for a semiflexible sequential copolymer chain and solved the partially directed
and fully directed self avoiding walk models of the copolymer chain for the adsorption
of the chain on an impenetrable surface perpendicular to the preffered direction of the walks of the polymer chain. 
Finally, in Sec. 3 we discuss the results obatined.

\section{Model and method}
A model of directed self-avoiding walk \cite{18} on a cubic lattice has been used to 
study adsorption-desorption phase transition behaviour of a sequential copolymer chain under 
good solvent condition. 
The directed walk models are
restrictive in the sense that anagle of bending has unique value that is $90^{\circ}$
and directedness of the walk amounts to some degree of stiffness in the 
copolymer chain because all
directions of the space are not treated equally. 
However, directed self avoiding walk model can be solved 
analytically and therefore gives the exact value of  
adsorption transition point of a semiflexible copolymer chain.  
We consider following two  
cases of directedness of copolymer chain in three dimensions:
In the case (i) partially directed self avoiding walk ($PDSAW$) model,
walker is allowed to walk along $\pm y$-direction, $+x$ and $+z$ 
directions while in the case (ii) fully directed self avoiding walk ($FDSAW$) 
model, the walker is allowed to take steps along $+x$, $+y$ and $+z$ directions only.

The stiffness in the sequential copolymer chain has
been introduced by associating an energy barrier for each bend in the walk of copolymer chain. 
The stiffness weight $k=exp(-\beta\epsilon_{b})$ where $\beta=(k_BT)^{-1}$ is 
inverse of the temperature and $\epsilon_b(>0)$ is the energy associated with 
each bend of the walk of copolymer chain. For $k=1$ or $\epsilon_{b}=0$ the copolymer chain is said to be flexible and for 
$0<k<1$ or $0<\epsilon_{b} <\infty$ the copolymer chain is said to be 
semiflexible. However, when $\epsilon_{b}\to\infty$ or $k\to0$, 
the copolymer chain has a rigid rod like shape.

The partition function of a semiflexible sequential copolymer chain made of two type of monomers (A \& B) can be written as, 
\begin{equation}
Z(x_1,x_2,k)={\sum}^{N=\infty}_{N=0}\sum_{ all\hspace{0.07cm}walks\hspace{0.07cm}of\hspace{0.05cm}N\hspace{0.05cm}steps} {x_1}^{N/2}{x_2}^{N/2}k^{N_b}
\end{equation}
where, $N_b$ is the total number of bends in a walk of $N$ steps (monomers),
$x_1$ and $x_2$ is the step fugacity of each monomer of A and B type monomers respectively.
For the sake of mathematical simplicity we have taken here onwards $x_1=x_2=x$. 

In the three dimensional directed walk model, one can consider two distinct surfaces; one parallel and the
other perpendicular to the directedness of the walks of the copolymer chain. 
In the case of homopolymer chain adsorption on an attractive impenetrable surface, 
it has been found that the features associated with the adsorption were
same in the case of both the surface orientaions of directed polymer chain and similarity is also found
in between isotropic and directed walk models of the homopolymer chain \cite{19}. However, critical value of surface
attraction for adsorption of isotropic chain is different than the directed polymer chain. 
Here, we report the results found 
using generating function method for the adsorption of directed semiflexible sequential copolymer chain on a surface 
perpendicular to the direction of directedness of the polymer chain in three dimensions.

In the case of three dimensional space, surface is of two dimensions i. e., a $x-y$ plane located at $z=0$. 
The walker is allowed to walk along $+x$, $\pm y$ and $+z$ directions in the three dimensional space.
The sequential copolymer chain is made of two type monomers, (A \& B). If first monomer (step) of the copolymer
chain (walk), which is grafted to the surface is of A-type then the component of partition function with initial step along $+x$ 
direction can be written as $S_{1x}$ and other component with first step along 
any one out of $\pm y$ directions as $S_{1y}$. Similary, if B-type monomer is the first monomer
of the copolymer chain and it is grafted on the surface. In this situation, if first monomer (step)
of the polymer chain (walk) is along $+x$ direction, the component of partition function is written as 
$S_{2x}$ and $S_{2y}$ is the component of partition function with first monomer of the copolymer 
chain which is B-type having first step along any one out of $\pm y$ directions. $Z$ is component of generating function perpendicular 
to the plane of surface.     

\subsection{\bf \it Adsorption of a semiflexible copolymer chain on a surface perpendicular 
to one out of the two preferred direction of the walk in three dimensions ($PDSAW$ model).}

Partition function of a surface interacting copolymer chain can be calculated using the
method of analysis discussed by Mishra {\it et al.}, \cite{19} and 
components of the partition function $Z_{PD-C}(k,\omega_1,\omega_2,x)$ of the copolymer chain interacting with
the surface having first monomer of A type can be written as follows:
\begin{equation}
\hspace{-2cm}S_{1x}(k,\omega_1,\omega_2,x)=s_1+\frac{s_1(s_2+2kS_{2y}+kZ)}{1-s_1s_2}+\frac{s_1s_2(s_1+2kS_{1y}+kZ)}{1-s_1s_2} \hspace{.6cm} (s_1s_2<1)
\end{equation}
where, $s_1(=\omega_1x)$ is the Boltzmann weight of interaction energy of A type monomer with the surface and 
similarly $s_2(=\omega_2x)$ is that of B type monomer.
\begin{equation}
\hspace{-2cm}S_{1y}(k,\omega_1,\omega_2,x)=s_1+\frac{s_1(s_2+2kS_{2x}+kZ)}{1-s_1s_2}+\frac{s_1s_2(s_1+2kS_{1x}+kZ)}{1-s_1s_2} \hspace{.6cm} (s_1s_2<1)
\end{equation}
and components of the partition function with first monomer of B type is,
\begin{equation}
\hspace{-2cm}S_{2x}(k,\omega_1,\omega_2,x)=s_2+\frac{s_2(s_1+2kS_{1y}+kZ)}{1-s_1s_2}+\frac{s_1s_2(s_2+2kS_{2y}+kZ)}{1-s_1s_2} \hspace{.6cm} (s_1s_2<1)
\end{equation}
\begin{equation}
\hspace{-2cm}S_{2y}(k,\omega_1,\omega_2,x)=s_1+\frac{s_2(s_1+2kS_{1x}+kZ)}{1-s_1s_2}+\frac{s_1s_2(s_2+2kS_{2x}+kZ)}{1-s_1s_2} \hspace{.6cm} (s_1s_2<1)
\end{equation}
while component perpendicular to the plane of the surface is \cite{19},
\begin{equation}
Z(k,x)=-\frac{x+(2k-1)x^2}{(-1-k+4k^2)x^2+(k+2)x-1}
\end{equation}
On solving Eqs. (2-5) and using value of $Z$ from Eq. (6), we get value of $S_{1x}$ and $S_{1y}$,
\begin{equation}
\hspace{-2.5cm}S_{1x}(k,\omega_1,\omega_2,x)=-\frac{s_1(-1+u_1s_2+u_2s_1{s_2}^2)(-1+2x+(-1+2k^2)x^2)}{(1-2s_1s_2(1+2k^2)+{s_1}^2{s_2}^2(1-2k^2)^2)u_3} \hspace{.5cm} (s_1s_2<1)
\end{equation}

where, $u_1$, $u_2$ and $u_3$ are:

$u_1=-1+s_1+2k^2s_1-2k(1+2s_1)$

$u_2=1-2k-2k^2+4k^3$

$u_3=(-1-k+4k^2)x^2+(k+2)x-1$

and

\begin{equation}
\hspace{-2.5cm}S_{1y}(k,\omega_1,\omega_2,x)=-\frac{(-s1+s_1s_2u_4+u_5s_1{s_2}^2)(-1+2x+(-1+2k^2)x^2)}{(1-2s_1s_2(1+2k^2)+{s_1}^2{s_2}^2(1-2k^2)^2)u_3} \hspace{.4cm} (s_1s_2<1)
\end{equation}

where,

$u_4=-1+s_1+2k^2s_1-k(1+2s_1)$

$u_5=1-k-2k^2+2k^3$

Thus, the partition function of the copolymer chain having first monomer of A type and grafted to the surface can be
written as,

\begin{equation}
Z_{PD-C}(k,\omega_1,\omega_2,x)=S_{1x}+2S_{1y}+Z \hspace{1cm} (s_1s_2<1)
\end{equation}

where,

\begin{equation}
\hspace{-2cm}Z_{PD-C}(k,\omega_1,\omega_2,x)=\frac{(u_6+u_7+{s_1}^2s_2(-24k^5s_2x^2+(-1+x)u_8+u_{9}))}{(1-2s_1s_2(1+2k^2)+{s_1}^2{s_2}^2(1-2k^2)^2)u_3} 
\end{equation}

here,

$u_6=x(-1+x-2kx)+s_1(-3+6x+(-3+6k^2)x^2$

$u_7=s_2(-3+8x-5x^2+16k^3x^2+2k^2x(2+x)+k(-4+8x))$

$u_8=-3-3s_2+3x+4s_2x$

$u_{9}=8k^3u_{10}+4k^4xu_{11}-2ku_{12}-2k^2u_{13}$ 

$u_{10}=s_2-2s_2x+2x^2+3s_2x^2$

$u_{11}=-3x+s_2(-1+4x)$

$u_{12}=4(-1+x)^2+s_2(2-4x+3x^2)$

and

$u_{13}=-3+6x+s_2(3-8x+8x^2)$

Singularities appearing in Eq. ($10$) give the critical value of 
$x_c=\frac{k+2-\sqrt{17}k}{2(1+k-4k^2)}$ \cite{19} and 
$\omega_{c1} = \frac{4(1+k-4k^2)^2}{(1+\sqrt2 k)^2(k+2-\sqrt{17} k)^2\omega_{c2}}$.
On substitution of $\omega_{c1}=\omega_{c2}=\omega_c$, we are able to obtain
$\omega_c$ required for adsorption of a semiflexible homopolymer
chain for $3D-PDSAW$, as reported by Mishra {\it et al.} \cite{19}.

We consider value of $\omega_{c2}$ equal to one, greater
than one (say, 1.5) and less than one (say, 0.5) depending on the fact that B-type monomer is having no interaction,
attractive or repulsive interaction with the surface and obtain $\omega_{c1}$ required for
adsorption of the copolymer chain on the surface. 
Variation of $\omega_{c1}$ is shown for different values of $\beta\epsilon_b$ for three values of
$\omega_{c2}$ in Fig. (1).

(ii) {\bf \it Fully directed self avoiding walk model:}

Partition function $Z_{FD-C}(k,\omega_1,\omega_2,x)$ for this
case can be easily evaluated following the method used for $3D-PDSAW$ model, discussed above. We write
components of the partition function $Z_{FD-C}(k,\omega_1,\omega_2,x)$ of a semiflexible
sequential copolymer chain having first monomer of A type as, 

\begin{equation}
\hspace{-2cm}S_{1x}(k,\omega_1,\omega_2,x)=s_1+\frac{s_1(s_2+kS_{2y}+kZ)}{1-s_1s_2}+\frac{s_1s_2(s_1+kS_{1y}+kZ)}{1-s_1s_2} \hspace{.6cm} (s_1s_2<1)
\end{equation}
\begin{equation}
\hspace{-2cm}S_{1y}(k,\omega_1,\omega_2,x)=s_1+\frac{s_1(s_2+kS_{2x}+kZ)}{1-s_1s_2}+\frac{s_1s_2(s_1+kS_{1x}+kZ)}{1-s_1s_2} \hspace{.6cm} (s_1s_2<1)
\end{equation}
and components of partition function with monomer grafted on the surface of B type are,
\begin{equation}
\hspace{-2cm}S_{2x}(k,\omega_1,\omega_2,x)=s_2+\frac{s_2(s_1+kS_{1y}+kZ)}{1-s_1s_2}+\frac{s_1s_2(s_2+kS_{2y}+kZ)}{1-s_1s_2} \hspace{.6cm} (s_1s_2<1)
\end{equation}
\begin{equation}
\hspace{-2cm}S_{2y}(k,\omega_1,\omega_2,x)=s_1+\frac{s_2(s_1+kS_{1x}+kZ)}{1-s_1s_2}+\frac{s_1s_2(s_2+kS_{2x}+kZ)}{1-s_1s_2} \hspace{.6cm} (s_1s_2<1)
\end{equation}
and component perpendicular to the surface is,
\begin{equation}
Z(k,x)=-\frac{x}{-1+x(1+2k)}
\end{equation}
On solving Eqs. (11-14) and substituting value of $Z$ from Eq. (15) we get following values of $S_{1x}$ and $S_{1y}$,
\begin{equation}
\hspace{-4.3cm}S_{1x}(k,\omega_1,\omega_2,x)=S_{1y}(k,\omega_1,\omega_2,x)=-\frac{s_1(1+s_2+ks_2)(-1+x+kx)}{(-1+{(1+k)}^2s_1s_2)(-1+x+2kx)} \hspace{1cm} (s_1s_2<1)
\end{equation}
\begin{equation}
Z_{FD-C}(k,\omega_1,\omega_2,x)= S_{1x}+S_{1y}+Z
\end{equation}
so that,
\begin{equation}
\hspace{-2cm}Z_{FD-C}(k,\omega_1,\omega_2,x)= \frac{x-s_1u_{14}}{(-1+{(1+k)}^2s_1s_2)(-1+x+2kx)} \hspace{1cm} (s_1s_2<1)
\end{equation}

where,

$u_{14}=2(-1+x+kx)+(1+k)s_2(-2+3(1+k)x)$
 
Singularities of the partition function, in this case  
give the critical value of $x_c = \frac{1}{2k+1}$ \cite{19} and $\omega_{c1} =\frac{(2k+1)^2}{\omega_{c2}(k+1)^2}$.
Assuming, $\omega_{c2}$ equal to one, greater
than one (say, 1.5, an attractive interaction of B-type monomers with the surface) 
and less than one (say, 0.5, a repulsive interaction of B-type monomers 
with the surface), we obtained $\omega_{c1}\ge1$ so that 
adsorption of copolymer chain may take place on the surface. 
Variation of $\omega_{c1}$ with bending energy of the copolymer chain is shown in Fig. 1 for
$FDSAW$ model on cubic lattice. In this case too, on substitution of $\omega_{c1}=\omega_{c2}=\omega_c$, 
we are able to reproduce
value of $\omega_c$ required for adsorption of the semiflexible homopolymer
chain, as reported by Mishra {\it et al.} \cite{19} for $FDSAW$ model on the cubic lattice.

\begin{figure}[htb] 
\hskip .1cm 
\centering 
\epsfxsize=14cm\epsfbox{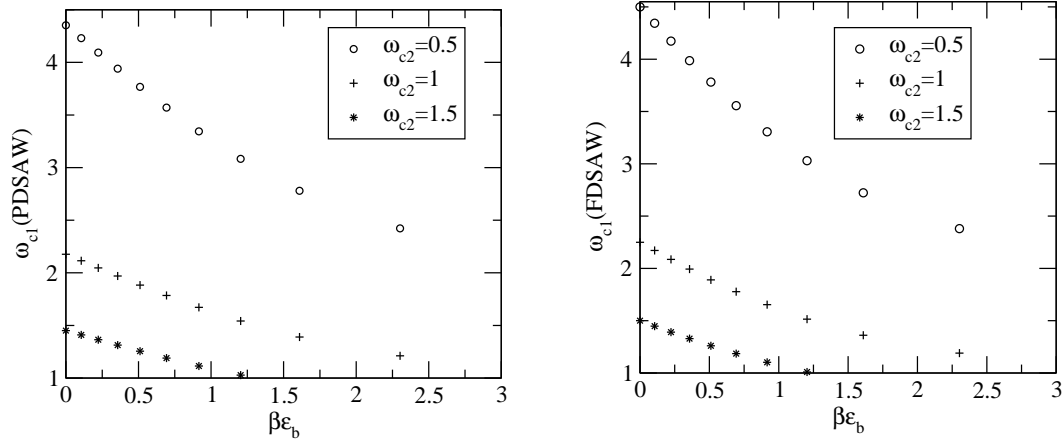} 
\caption{This figure compares the values of $\omega_{c1}$ for different values of 
$\beta \epsilon_b$ for adsorption of a semiflexible sequential copolymer chain on a surface perpendicular to one of the  
preferred directions of the copolymer chain for $PDSAW$ and $FDSAW$ models for three values of $\omega_{c2}=0.5, 1 \& 1.5$.}  
\end{figure} 

\section{Result and discussion}
Directed self avoiding walk model for a semiflexible sequential copolymer chain has been solved analytically in three
dimension under good solvent condition. 
For three dimensional space the surface considered is a two dimensional 
impenetrable plane
and the copolymer chain is made of two type of monomers (A \& B) and walk of the copolymer chain is
directed perpendicular to the plane of surface. The monomers of the copolymer chain
are interacting with surface. Using generating function technique,  
we have obtained critical value of monomer surface 
attraction required for adsorption of the directed copolymer chain on the surface. The 
dependence of the critical value of surface attraction on the stiffness of 
the copolymer chain has been evaluated for the cases in which one type of monomer of the
copolymer chain is having attractive interaction,
repulsive interaction or no interaction with the surface. Our study showed that adsorption of a stiffer 
copolymer chain takes place at smaller value of monomer surface attraction 
when compared to that of a flexible copolymer
chain. These features are similar to the semiflexible homopolymer adsorption \cite{19}. 

The 
observed feature that stiff chain adsorption occurs at a smaller value of monomer surface attraction than the flexible chain 
can be understood on physical ground in following manner. The 
adsorption desorption phase transition behaviour of the polymer chain is result of competition between the 
gain in internal
energy to the polymer chain due to surface attraction and loss of the entropy of the chain due to presence
of the surface close to the chain. Stiffer chain has less entropy than flexible polymer chain, therefore, 
entropy loss is small when stiffer chain adsorbs onto the surface than the flexible chain. That is why the stiff
chain adsorbs onto the surface at a smaller value of monomer surface attraction than the flexible polymer chain.

In this calculation we have considered step fugacity of both type of monomers same, i. e. $x_1=x_2=x$.
However, this calculation can also be done using method of analysis
discussed here for the case $x_1\ne x_2$.  
The details of mathematical parts of this work are planned to
be published elsewhere.

\end{document}